\documentclass[aip,amsmath,amssymb,seceqn]{revtex4-1}
\usepackage{graphicx}
\usepackage{graphics}
\usepackage{epsfig}
\usepackage{dcolumn}
\usepackage{bm}
\usepackage{makeidx}
\usepackage{subfigure}
\usepackage{color}
\usepackage{longtable}
\usepackage{lineno}

\begin{document}

\title{Antiferromagnetic multi-level memory cell}
\author{V. Schuler$^{\ast}$}
\author{K. Olejn\'{\i}k$^{\ast,\dagger}$}
\author{X.~Marti}
\author{V. Nov\'ak}
\affiliation{Institute of Physics, Academy of Sciences of the Czech Republic, Cukrovarnick\'a 10, 162 00 Praha 6, Czech Republic}
\author{P.~Wadley}
\author{R.~P.~Campion}
\author{K.~W.~Edmonds}
\author{B.~L.~Gallagher}
\affiliation{School of Physics and Astronomy, University of Nottingham, Nottingham NG7 2RD, United Kingdom}
\author{J.~Garces}
\affiliation{IGS Research, Calle La Coma, Nave 8, 43140 La Pobla de Mafumet, Tarragona, Spain}
\author{M.~Baumgartner}
\author{P.~Gambardella}
\affiliation{Department of Materials, ETH Z\"urich, H\"onggerbergring 64, CH-8093 Z\"urich, Switzerland}
\author{T. Jungwirth}
\affiliation{Institute of Physics, Academy of Sciences of the Czech Republic, Cukrovarnick\'a 10, 162 00 Praha 6, Czech Republic}
\affiliation{School of Physics and Astronomy, University of Nottingham, Nottingham NG7 2RD, United Kingdom}


\begin{abstract}
$^\ast$These authors contributed equally to the work

\noindent $^\dagger$To whom correspondence should be addressed; E-mail: olejnik@fzu.cz 
\end{abstract}

\maketitle

{\bf
Antiferromagnets (AFs) are remarkable magnetically ordered materials that due to the absence of a net magnetic moment do not generate dipolar fields and are insensitive to external magnetic field perturbations. However, it has been notoriously difficult to control  antiferromagnetic moments by any practical means suitable for device applications. This has left AFs over their hundred years history\cite{Neel1970} virtually unexploited and only poorly explored, in striking contrast to the thousands of years of fascination and utility of ferromagnetism. Very recently it has been predicted and experimentally confirmed that relativistic spin-orbit torques can provide the means for efficient electrical control of an AF.\cite{Zelezny2014,Wadley2016}  Here we place the emerging field of antiferromagnetic spintronics on the map of non-volatile solid state memory technologies.  We demonstrate the complete write/store/read functionality in an antiferromagnetic CuMnAs bit cell embedded in a standard printed circuit board communicating with a computer via a USB interface. We show that the elementary-shape bit cells fabricated from a single-layer AF are electrically written on timescales ranging from milliseconds to nanoseconds  and we demonstrate their deterministic multi-level switching. The multi-level cell characteristics, reflecting series of reproducible, electrically controlled domain reconfigurations, allow us to integrate memory and signal counter functionalities within the bit cell.
}

In ferromagnetic materials, all the magnetic moments sitting on individual atoms point in the same direction and can be switched by running an electrical current through a nearby electromagnet. This is the principle of recording in ferromagnetic media used from the 19th century magnetic wire recorders till today's hard-drives.   Magnetic storage has remained viable throughout its entire history and today is the key technology providing the virtually unlimited data space on the internet. To keep it viable, the 19th century inductive coils were first removed from the readout and replaced by the 20th century spin-based magneto-resistive technology.\cite{Chappert2007}  21st century physics brought yet another revolution by eliminating the electromagnetic induction from the writing process in magnetic memory chips  and replacing it with the spin-torque phenomenon.\cite{Chappert2007} In the non-relativistic version of the effect, switching of the recording ferromagnet is achieved by electrically transferring spins from a fixed reference permanent magnet. In the recently discovered relativistic version of the spin torque,\cite{Chernyshov2009,Miron2011b,Liu2012,Garello2013}  the reference magnet is eliminated and the switching is triggered by the internal transfer from the linear momentum to the spin angular momentum under the applied writing current.\cite{Sinova2015}  The complete absence of electromagnets or reference permanent magnets in this most advanced physical scheme for writing in ferromagnetic spintronics has served as the key for introducing the new physical concept\cite{Zelezny2014} for the efficient control of magnetic moments in AFs that underpins our work.

In their simplest form, compensated AFs have north poles of half of the microscopic atomic moments pointing in one direction and the other half in the opposite direction. This makes the external magnetic field inefficient for switching magnetic moments in AFs. Instead, our devices rely on the recently dicovered special form of the relativistic spin torque.\cite{Zelezny2014,Wadley2016}  When driving a macroscopic electrical current through certain antiferromagnetic crystals whose magnetic atoms occupy inversion-partner lattice sites (e.g. in AF CuMnAs or Mn$_2$Au), a local relativistic field is generated which points in the opposite direction on magnetic atoms with opposite magnetic moments. The staggered relativistic field is then as efficient in switching the AF as a conventional uniform magnetic field in switching a ferromagnet. This reverses the traditionally skeptical perception of the utility of AFs in microelectronics. Simultaneously, AFs offer a unique combination of radiation and magnetic field hardness, internal spin dynamics frequency-scales reaching THz, and the absence of dipolar magnetic fields. This opens new avenues for spintronics research and applications,\cite{MacDonald2011,Gomonay2014,Jungwirth2016,Baltz2016} including the realization of solid state memories we focus on in this work.  

Fig.~1 provides an overview of the basic characteristics of our antiferromagnetic CuMnAs memory cells. For the purpose of the present study the cell has a cross shape, 2$\mu$m in size, patterned by electron beam lithography and reactive ion etching from a single-crystal CuMnAs film grown by molecular beam epitaxy\cite{Wadley2013} (Fig. 1a). Writing current pulses, depicted by red arrows in Fig. 1c, are sent through the four contacts to generate current lines in the central region of the cross along either the [100] or [010] CuMnAs crystal axis. The writing current pulses give preference to domains with antiferromagnetic moments aligned perpendicular to the current lines,\cite{Zelezny2014,Wadley2016} as shown schematically in Fig. 1c by the white double-arrows. Electrical readout is done by running the probe current along one of the arms of the cross (blue arrow in Fig. 1c) and by measuring the antiferromagnetic transverse anisotropic magnetoresistance (planar Hall effect) across the other arm.\cite{Wadley2016,Kriegner2016}  

The cell input and output signals can be sent at ambient conditions using a printed circuit board containing the antiferromagnetic memory chip and other standard components and connected to a personal computer via a USB interface (Fig. 1b). Examples of different write-pulse sequences and corresponding readout signals obtained with this demonstrator USB device are shown in Figs.~1d,e. In one case a symmetric pulsing was applied, repeating four pulses with current lines along the [100] direction followed by four pulses with current lines along the [010] direction. In a second case, the four pulses with current lines along the [100] direction are followed by fifty pulses with current lines along the [010] direction. The results illustrate a deterministic multi-level switching of the CuMnAs bit cell. A complementary  photoemission electron microscopy study of CuMnAs has associated the electrical switching signal with the antiferromagnetic moment reorientations within multiple domains of sub-100~nm dimension.\cite{Grzybowski2016}  Figs.~1d,e highlight the level of electrical control that can be achieved over these multi-domain switching processes in antiferromagnets which, unlike ferromagnets,\cite{Lequeux2016} are insensitive to and do not generate dipolar magnetic fields.

We now proceed to exploring in detail the dependencies of the readout signals on the writing characteristics, namely on the pulse length and amplitude, on the number of pulses and duty cycle. Since our study covers a broad range of parameters we performed the experiments using laboratory electrical pulse generators or high frequency set-ups equipped with rf cables and the AF devices mounted on specially designed co-planar waveguide with rf access. 

In Fig. 2a we plot a typical dependence of the measured readout signal on the length of a single writing pulse. Before each measurement of the given pulse length, the cell was reset to the same initial state and then the single write pulse was applied with the current lines along the [100] direction. The initial linear increase of the readout signal with increasing pulse length defines the signal per pulse length ratio which we plot in Fig.~2b as a function of the writing current density. For comparison, we included in the plot also data points for a 30~$\mu$m-size cell used in earlier measurements reported in Ref.~\onlinecite{Wadley2016}. The larger cell allowed us to explore only a  limited range of current densities before heating damaged the sample. On the other hand, as shown in Fig.~2c, these larger devices (prepared by optical lithography and wet etching) also show the deterministic multi-level characteristics as the 2~$\mu$m cells (cf. Fig. 1). 
For the 2~$\mu$m cells, much high current densities can be applied which allowed us to scale  the writing pulse length from milliseconds\cite{Wadley2016} down to nanoseconds while keeping the m$\Omega$ level of the readout signal, as illustrated in Fig. 2d. Remarkably, increasing the current density by only a factor of 4 in the 2~$\mu$m cell was sufficient for enhancing the signal per pulse length ratio by six orders of magnitude (see Fig. 2b) desired for observing the nanosecond-pulse switching.  

We note that the elementary design of our CuMnAs bit cell leaves still a room for a significant downscaling of its size which is necessary for advancing to the picosecond limit of switching times accessible in AFs.\cite{Kimel2004} Electrical pulses of this ultra-short length have to be, however, triggered optically which is beyond the scope of the present work. Instead, we focus in the remaining paragraphs on the multi-level bit cell characteristics when written by trains of pulses with the individual pulse length varied from milliseconds to microseconds. The results, summarized in Figs. 3a-d, were obtained using the following measurement protocol: Before each train of pulses (with writing current lines along the [100] direction), the cell was reset to the same initial state. The maximum length of the pulse train, including all pulses and delays between pulses, was set to 100~ms and readout was performed 5~s after the last pulse in the train. The current density was fixed at 2.7$\times 10^7$~Acm$^{-2}$.

In Fig.~3a we compare the dependencies of the readout signal on the number of pulses for different individual pulse lengths. The dependencies are highly reproducible as indicated by error bars obtained from repeated measurements for each pulse train. The antiferromagnetic bit cell acts as a counter of pulses whose number can be in hundreds. The separation of the readout signals for different numbers of pulses, i.e. the accuracy of the pulse counting, increases with increasing individual pulse length and can reach a single-pulse resolution. The duty cycle was fixed in all measurements shown in Fig.~3a to 0.025. In Fig.~3b we show that for a given individual pulse length, the duty cycle (delay between pulses) can be varied over a broad range without affecting the readout signal of the counter.

In Figs.~3c,d we plot the readout signal dependence on the integrated pulse time, i.e., on the number of pulses multiplied by the individual pulse length. Over a broad range of individual pulse lengths, the dependencies fall onto a universal curve making the antiferromagnetic memory cell a detector of the integrated pulse time, as shown in Fig.~3c. The universal trend breaks down for individual pulse lengths smaller than $\approx50$~$\mu$s. As already indicated by the comparison between larger and smaller cells in Fig.~2b, heating assists the spin-orbit torque switching in our devices which can also explain the trends seen in Fig.~3c. By a direct measurement of the heating during the pulse we observe that in the 2~$\mu$m cells the heating saturates at pulse lengths exceeding 10's of $\mu$s. For these longer pulses, switching occurs at the saturated temperature which results in the universal dependence of the readout signal on the integrated pulse time. For shorter pulses, the temperature during switching does not reach saturation and the heating decreases with decreasing pulse length which results in the lower readout signal. We note that in all measurements the temperature during switching stays at least 100 degrees below the CuMnAs N\'eel temparture ($T_N=480$~K)\cite{Wadley2015a}. 

An accurate detection of the integrated pulse time is feasible for 10's of pulses in our measurements, as shown in Fig.~3d. For pulse numbers exceeding 100's, the readout signal at a given integrated pulse time drops down from the universal trend because of the non-saturated heating during the shorter pulses. The signal reduction gets stronger at lower integrated pulse times with correspondingly smaller individual pulse lengths. For the current density used in the measurements in Fig.~3, the readout signal vanishes for pulse lengths below a microsecond.

The deterministic multi-level memory characteristics described above have been consistently observed in bit cells fabricated in our single-layer antiferromagnetic CuMnAs deposited directly on a III-V substrate. The cells have an elementary cross-shape geometry with no intentional complexities introduced either during the layer growth or device fabrication. Antiferromagnets thus appear as particularly favorable materials for realizing multi-level memories.\cite{Kriegner2016,Fukami2016} Moreover, by utilizing the current-induced spin-orbit torques,  the multi-level antiferromagnetic bit cells are compatible with common electronic circuitry and can combine memory  with additional functionalities. These may include, as demonstrated above,  pulse number or pulse time counters. We anticipate that a range of other effects and functionalities will emerge with more elaborate materials and device engineering, and with switching times scaled from the presently demonstrated nanoseconds down by another few orders of magnitude  before reaching the physical spin dynamics limit in antiferromagnets. This opens a new research avenue with information technologies at its horizon based on an entirely new type of components -- multi-functional non-volatile antiferromagnetic devices, breaking present limits on speed, energy efficiency, integration, and security. The emerging or future technology areas may include antiferromagnetic multi-level memory cells for elementary signal processing and data storage in human-interface devices and internet of things applications, radiation and magnetic-field hard, non-volatile computer memories, or unrivalled, ultra-fast functional memories for future photonic information technologies.

\section*{Methods}
\subsection*{CuMnAs growth}
The CuMnAs film of 60~nm thickness  was grown on GaP(001) by molecular beam epitaxy at a substrate temperature of 300$^\circ$C. X-ray diffraction measurements showed that the film has the tetragonal Cu$_2$Sb structure (space group P4/nmm). The film and substrate follow the epitaxial relationship CuMnAs(001)[100]$\parallel$GaP(001)[110], with less than 1\% lattice mismatch. Magnetic measurements confirmed that the CuMnAs film is a compensated antiferromagnet. The N\'eel temperature is 480~K.
\subsection*{Lithography}
Microfabrication of our CuMnAs cross-shape cells was done using electron beam lithography and reactive ion etching. We used metal masks prepared by lift-off method directly on the surface of CuMnAs films to define the pattern and protect CuMnAs layer during the reactive ion etching step. We have developed fabrication recipes enabling us to control the degree of oxidation of the CuMnAs surface below contact pads that is necessary to control the overall resistance of the devices.
\subsection*{USB printed circuit board}
The electronic board for operating the antiferromagnetic memory cell is powered by a (5~V) USB 2.0 socket and contains a standard microcontroller chip and transistors for opening connections to the individual arms of the cross-shape cell. Readout signal is processed in a standard chip containing an amplifier ($\times$ 8) and a 16-bit analogue-digital converter operating in the range $\pm2$~V after the amplification stage. Communication with PC is performed using the USB socket operating as a virtual serial port.

\section*{Acknowledgements}
We acknowledge support from the EU ERC Advanced Grant No. 268066, from the Ministry of Education of the Czech Republic Grant No. LM2011026, from the Grant Agency of the Czech Republic Grant no. 14-37427, from the University of Nottingham EPSRC Impact Acceleration Account grant No. EP/K503800/1, and from the Swiss National Science Foundation, Grant No. 200021-153404.
\section*{Author contributions}

\bibliography{Refs}

\protect\newpage

\begin{figure}[h!]
\hspace*{-0cm}\epsfig{width=.8\columnwidth,angle=0,file=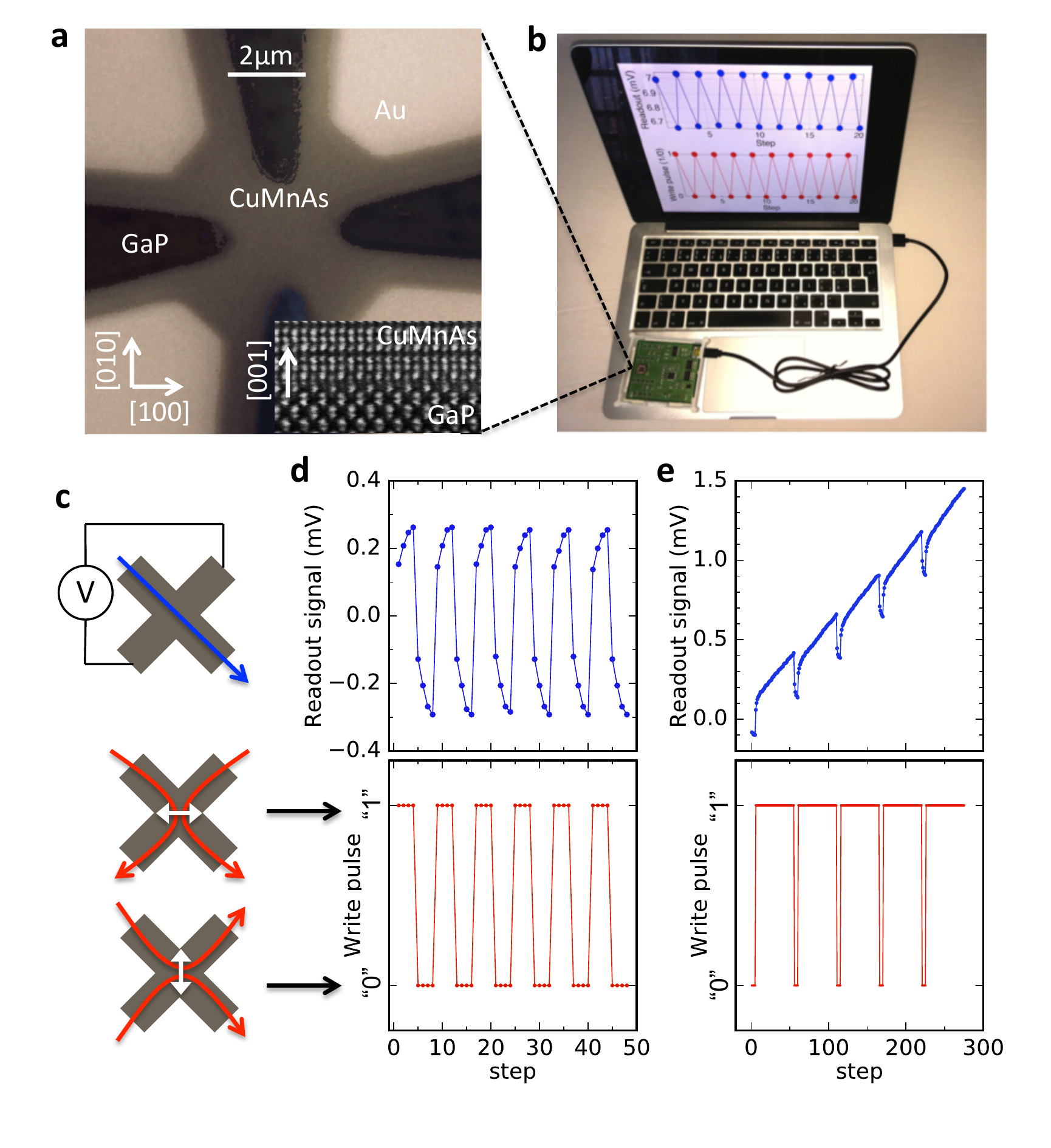}
\caption{{\bf a}, Optical microscopy image of the device containing Au contact pads (light) and the antiferromagnetic CuMnAs cross-shape cell on the GaP substrate (dark). Inset: Scanning transmission electron microscopy image of CuMnAs/GaP in the [100]-[001] plane. {\bf b}, Picture of the printed circuit board with the antiferromagnetic chip and the input write-pulse signals (red dots) and output readout signals (blue dots) send via a USB computer interface. {\bf c}, Top: The readout current (blue arrow) and transverse voltage detection geometry. Middle and bottom: Write pulse current lines (red arrows) labeled "1" (middle) and "0" (bottom) and the corresponding preferred antiferromagnetic moment orientations (white double-arrows). {\bf d}, Bottom: A symmetric pulsing with repeated four write pulses with current lines along the [100] direction ("0") followed by four pulses with current lines along the [010] direction ("1"). Top: corresponding readout signals. {\bf e}, Same as  {\bf d} with the four "0" write pulses followed by fifty "1" pulses.}
\label{fig1}
\end{figure}

\begin{figure}[h!]
\hspace*{-0cm}\epsfig{width=1\columnwidth,angle=0,file=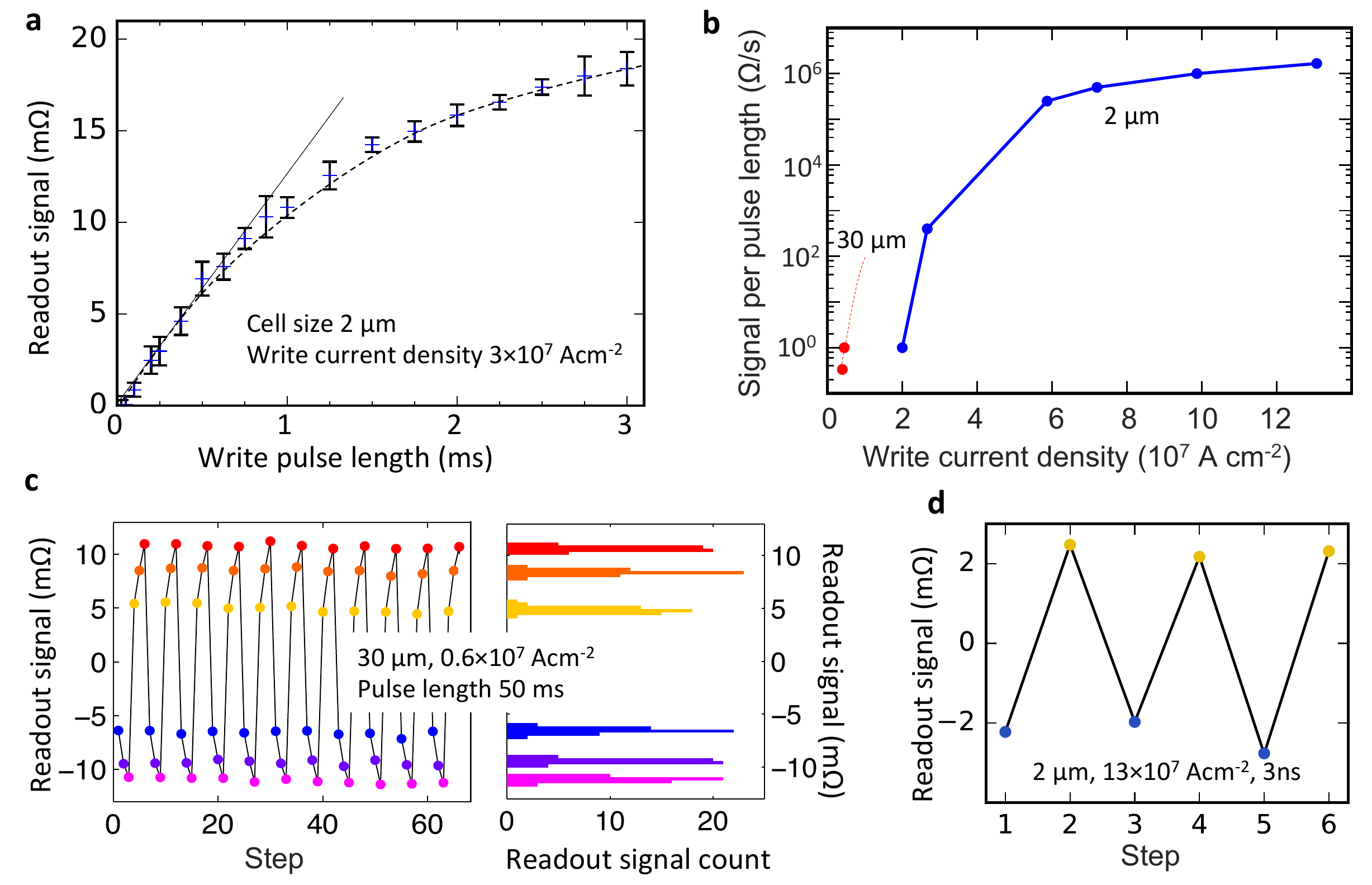}
\caption{{\bf a}, Readout signal of a 2~$\mu$m device as a function of the applied write-pulse length at a fixed current density of $3\times10^7$~Acm$^{-2}$ and current lines along the [100] direction. All data points are obtained starting from the same reference state. Reading is performed with a current density of $5\times10^5$~Acm$^{-2}$, 5~s after the write pulse. The initial linear slope of the dependence (signal per pulse length ratio) is highlighted by a linear fit. {\bf b}, Readout signal per write-pulse length rate obtained from the initial linear slope (see panel {\bf a}) as a function of the write current density, for a 30~$\mu$m device (red) and 2~$\mu$m device (blue). {\bf c}, Multi-level switching in the 30~$\mu$m device. Three pulses are applied along the [100] direction followed by three pulses along the [010] direction with current density of $0.6\times10^7$~Acm$^{-2}$ and pulse length 50~ms. Right: Histogram of the six different states, obtained from 50 repetitions of the 3+3 pulse sequence. {\bf d}, Switching measurement in the 2~$\mu$m device when repeating a single 3~ns pulse (blue dots) of current density $13\times10^7$~Acm$^{-2}$ along the [100] direction followed by a single 50~ms pulse of current density $2\times10^7$~Acm$^{-2}$ along the [010] direction. (High frequency rf access was available only for the [100] switching current direction.) All measurements were performed at room temperature.}
\label{fig2}
\end{figure}

\begin{figure}[h!]
\hspace*{-0cm}\epsfig{width=1\columnwidth,angle=0,file=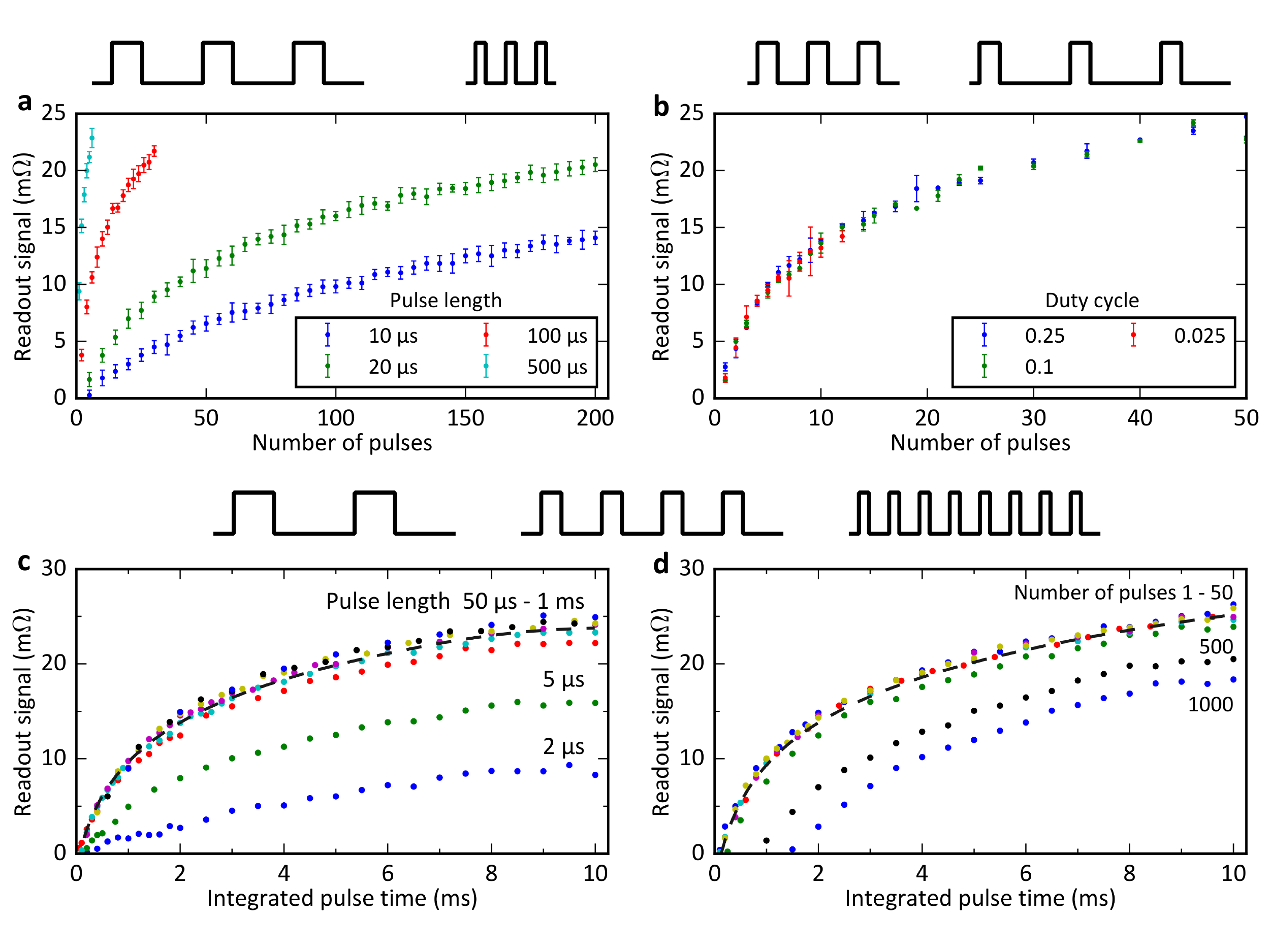}
\caption{{\bf a}, Readout signal as a function of the number of pulses in the train of pulses, for different values of the individual pulse length and a common duty cycle of 0.025. All data points are obtained starting from the same reference state. The writing current density is $2.7\times10^7$~Acm$^{-2}$.  Plotted data points are the average over ten measurements; error bars represent the standard deviation. {\bf b}, Same as {\bf a} for different duty cycles (corresponding to different delays between individual pulses) and for a common individual pulse length of 200~$\mu$s. {\bf c}, Same as {\bf a} measured as a function  of the integrated pulse time and plotted for different individual pulse lengths. {\bf d}, Same as {\bf c}  plotted for different number of pulses in the pulse train.}
\label{fig3}
\end{figure}

\end{document}